\documentclass[a4paper,fleqn,usenatbib]{mnras}

\usepackage{newtxtext,newtxmath}

\usepackage[T1]{fontenc}
\usepackage{ae,aecompl}

\usepackage{graphicx}	
\usepackage{amsmath}	
\usepackage{amssymb}	
\usepackage{times}
\usepackage{graphics, subfigure}
\usepackage{epsfig}
\usepackage{multirow}
\usepackage{ulem}
\usepackage{pdfpages}

\def\simlt{\mathrel{\hbox{\rlap{\hbox{\lower4pt\hbox{$\sim$}}}\hbox{$<$}}}}
\def\simgt{\mathrel{\hbox{\rlap{\hbox{\lower4pt\hbox{$\sim$}}}\hbox{$>$}}}}

\newcommand{\mysim}{\mathord{\sim}}

\title[comments on numerical detonations]{Comments on "Numerical Stability of Detonations in White Dwarf Simulations"}

\author[D. Kushnir and B. Katz]{
Doron Kushnir$^{1}$\thanks{E-mail: doron.kushnir@weizmann.ac.il} and Boaz Katz$^{1}$
\\
$^{1}$Dept. of Particle Phys. \& Astrophys., Weizmann Institute of
Science, Rehovot 76100, Israel
}

\date{Accepted XXX. Received YYY; in original form ZZZ}

\pubyear{2019}

\begin{document}
\label{firstpage}
\pagerange{\pageref{firstpage}--\pageref{lastpage}}
\maketitle

\begin{abstract}
Katz \& Zingale (2019, KZ19) recently studied a one-dimensional test problem, intended to mimic the process of detonation ignition in head-on collisions of two carbon--oxygen (CO) white dwarfs. They do not obtain ignition of a detonation in pure CO compositions unless the temperature is artificially increased or $5\%$ He is included. In both of these cases they obtain converged ignition only for spatial resolutions better than $0.1\,\textrm{km}$, which are beyond the capability of multidimensional simulations. This is in a contradiction with the claims of Kushnir et al. (2013, K13), that a convergence to $\mysim10\%$ is achieved for a resolution of a few km. Using Eulerian and Lagrangian codes we show that a converged and resolved ignition is obtained for pure CO in this test problem without the need for He or increasing the temperature. The two codes agree to within $1\%$ and convergence is obtained at resolutions of several km. We calculate the case that includes He and obtain a similar slow convergence, but find that it is due to a boundary numerical artifact that can (and should) be avoided. Correcting the boundary conditions allows convergence with resolution of  $\sim10\,\textrm{km}$ in an agreement with the claims of K13. It is likely that the slow convergence obtained by KZ19 in this case is because of a similar boundary numerical artifact, but we are unable to verify this. KZ19 further recommended to avoid the use of the burning limiter introduced by K13. We show that their recommendation is not justified. \end{abstract}

\begin{keywords}
hydrodynamics -- shock waves -- supernovae: general 
\end{keywords}



\section{Introduction}
\label{sec:Introduction}

It is widely believed that Type Ia supernovae result from the explosions of white dwarfs (WDs) composed predominantly of carbon and oxygen (CO), but there is no consensus regarding the explosion mechanism \citep[see][for a review]{Maoz2014}. A serious concern for many scenarios is that a successful ignition of an explosive detonation has never been convincingly demonstrated, which have led to the commonly introduced free parameters such as the deflagration velocity or transition to detonation criteria \citep[e.g., in the single-degenerate and double-degenerate scenarios; see][]{Hillebrandt2000}. The situation is different for direct collisions of CO WDs, where the nuclear detonations in these collisions are due to a well-understood shock ignition \citep{Kushnir2013}.

\citet{Katz2019} recently studied a one-dimensional test problem, intended to mimic the process of detonation ignition in WDs collisions. The test problem consists of an infinite slab with a uniform density colliding into a rigid wall with some initial velocity and a constant (in time and space) acceleration. This is a simplified version of a similar setup suggested by \citet{Kushnir2013}, where a non-uniform density profile equal to the density profile of a CO WD (that decreases to zero at the rigid wall) was used.

\citet{Katz2019} claimed that for their test problem with CO (equal mass fractions) composition, ignition is not reached unless the temperature is artificially increased at the onset of the simulation. They obtained ignition in a mixed composition with significant fraction of He (HeCO, 5\% He, 50\% C,  45\% O), but claimed to obtain numerical ignition near the boundary that disappears at higher resolutions, and allows converged ignition only for spatial resolutions better than $0.1\,\textrm{km}$. This is in a contradiction with the results of \citet{Kushnir2013}, that obtained converged ignition in pure CO with a resolution of a few km. 

Following the description of the simulation details in Section~\ref{sec:details}, we show in Section~\ref{sec:CO results} that resolved CO ignition occurs in the test problem employed by \citet{Katz2019}, by using Lagrangian and Eulerian codes (that solve the hydrodynamic equations and couple the burning to the hydrodynamics in a completely different way). Convergence of the ignition location to an accuracy better than $10\%$ ($1\%$) is obtained with a resolution of $100\,\textrm{km}$ ($10\,\textrm{km}$). We show that the ignition region is converged and has a width of $\mysim100\,\textrm{km}$. \citet{Katz2019} failed to obtain an ignition of a detonation, because they arbitrarily stopped the simulation after $3.5\,\textrm{s}$\footnote{M. Katz, private communication.}, roughly $0.2\,\textrm{s}$ before ignition is obtained in our simulations\footnote{Also verified by M. Katz with their numerical code.}. 

In Section~\ref{sec:HeCO results}, we reproduce the results of \citet{Katz2019} for HeCO composition. In our simulations, the numerical ignition near the boundary is a result of a numerical instability in the first few cells, which is due to the boundary conditions employed, and is enhanced by the fast reaction of $^{16}$O$+\alpha$. It is straightforward to avoid this numerical instability, by using custom appropriate boundary conditions, which allows convergence with a resolution of $\mysim10\,\textrm{km}$. It is likely that the slow convergence obtained by \citet{Katz2019} in this case is because of a similar boundary numerical artifact, but we are unable to verify this.

In Section~\ref{sec:discussion}, we summarize the results. In particular we address claims by \citet{Katz2019} regarding the use of a burning limiter introduced by \citet{Kushnir2013}. 

We note that the simplified test problem considered by \citet{Katz2019} and here, which employs a uniform density, misses important dynamical effects of WD collisions that are captured by the more realistic one-dimensional setup used in \citet{Kushnir2013}, such as the diminishing of the density and the divergence of the speed of sound towards the rigid wall, in regions where the ignition takes place, following the collision \citep[see detailed discussion in][]{Kushnir2014}. Also, the time-scale to ignition and the ignition distance from the rigid wall are larger in this setup (by factors of a few) than those obtained in the corresponding WDs collisions. Therefore, the results of this paper cannot be used directly to support the ignition that follows a collision of CO WDs \citep[as was demonstrated by][]{Kushnir2013}. Nevertheless, the simpler uniform-density setup captures important aspects of the ignition process, such as the size of the hotspot and the ability of the numerical schemes to resolve it, and we use it here to allow direct comparison with the results of \citet{Katz2019}. 

In what follows we normalize temperatures, $T_{9}=T[\textrm{K}]/10^{9}$. Some aspects of this work were performed with a modified version of the {\sc MESA} code\footnote{Version r7624; https://sourceforge.net/projects/mesa/files/releases/} \citep{Paxton2011,Paxton2013,Paxton2015}.


\section{Simulation details}
\label{sec:details}

In this section we describe the problem setup (Section~\ref{sec:setup}), the input physics for our simulations (Section~\ref{sec:input}) and we provide some details regarding the two numerical codes that we use (Section~\ref{sec:numerical codes}).

\subsection{Problem setup}
\label{sec:setup}

The one-dimensional test problem of \citet{Katz2019} consists of an infinite slab (occupying the region $x>0$) with a uniform density of $5\times10^{6}\,\textrm{g}\,\textrm{cm}^{-3}$ and temperature of $T_{0,9}=0.01$ colliding into a rigid wall at $x=0$ with a velocity of $-2\times10^{8}\,\textrm{cm}\,\textrm{s}^{-1}$ and a uniform, constant, acceleration\footnote{Note that \citet{Katz2019} reported a velocity of $-2\times10^{8}\,\textrm{m}\,\textrm{s}^{-1}$ and acceleration of  $-1.1\times10^{8}\,\textrm{m}\,\textrm{s}^{-2}$, but this is a typo (M. Katz, private communication)}. of $-1.1\times10^{8}\,\textrm{cm}\,\textrm{s}^{-2}$.

\subsection{Input physics}
\label{sec:input}

Our input physics, which we briefly summarize below, are very similar to the ones used by \citet{Kushnir2019}. More details can be found 
in \citet{Kushnir2019}.

We use the NSE$5$ list of $179$ isotopes \citep{Kushnir2019} without $^{6}$He, such that our list is composed of $178$ isotopes. For this list, the burning scales are converged to better than a percent \citep{Kushnir2019}. The nuclear masses were taken from the file \textsc{winvn\_v2.0.dat}, which is available through the JINA reaclib data base\footnote{http://jinaweb.org/reaclib/db/} \citep[JINA,][]{Cyburt2010}. For the partition functions, $w_{i}(T)$, we use the fit of \citet{Kushnir2019} for the values that are provided in the file \textsc{winvn\_v2.0.dat} over some specified temperature grid. 

The forward reaction rates are taken from JINA (the default library of 2017 October 20). All strong reactions that connect between isotopes from the list are included. Inverse reaction rates were determined according to a detailed balance. Enhancement of the reaction rates due to screening corrections is described at the end of this section. We further normalized all the channels of the $^{12}$C+$^{16}$O and $^{16}$O+$^{16}$O reactions such that the total cross-sections are identical to the ones provided by \citet{CF88}, while keeping the branching ratios provided by JINA. 

The equation of state (EOS) is composed of contributions from electron--positron plasma, radiation, ideal gas for the nuclei, ion--ion Coulomb corrections, and nuclear level excitations. We use the EOS provided by {\sc MESA} for the electron--positron plasma, for the ideal gas part of the nuclei, for the radiation and for the Coulomb corrections (but based on \citet{Chabrier1998} and not on \citet{Yakovlev1989}, see below). The electron--positron part is based on the \textit{Helmholtz} EOS \citep{Timmes00}, which is a table interpolation of the Helmholtz free energy as calculated by the Timmes EOS \citep{Timmes1999} over a density--temperature grid with $20$ points per decade. This is different from \citet{Kushnir2019}, where the Timmes EOS was used for the electron--positron plasma, since the \textit{Helmholtz} EOS is more efficient and because the internal inconsistency of the \textit{Helmholtz} EOS \citep[see][for details]{Kushnir2019} is small enough within the regions of the parameter space studied here. We further include the nuclear level excitation energy of the ions, by using the $w_{i}(T)$ from above.

We assume that the Coulomb correction to the chemical potential of each ion is given by $\mu_{i}^{C}=k_{B}Tf(\Gamma_{i})$ and is independent of the other ions \citep[linear mixing rule (LMR),][]{Hansen1977}, where $k_{B}$ is Boltzmann's constant,  $\Gamma_{i}=Z_{i}^{5/3}\Gamma_{e}$ is the ion coupling parameter and $\Gamma_{e}\approx(4\upi\rho N_{A} Y_{e}/3)^{1/3}e^{2}/k_{B}T$ is the electron coupling parameter. We use the three-parameter fit of \citet{Chabrier1998} for $f(\Gamma)$.  Following \citet[][]{Khokhlov88}, we approximate the LMR correction to the EOS by $f(\Gamma)$ for a `mean' nucleus $\Gamma=\bar{Z}^{5/3}\Gamma_{e}$. The screening factor for a thermonuclear reaction with reactants $i=1,..,N$ and charges $Z_{i}$ is determined from detailed balance \citep{KushnirScreen}:
\begin{eqnarray}\label{eq:NSE screening}
\exp\left(\frac{\sum_{i=1}^{N}\mu_{i}^{C}-\mu_{j}^{C}}{k_{B}T}\right),
\end{eqnarray}
where isotope $j$ has a charge $Z_{j}=\sum_{i=1}^{N}Z_{i}$ \citep[same as equation~(15) of][for the case of $N=2$]{Dewitt1973}.  

\subsection{Numerical codes}
\label{sec:numerical codes}

We provide below some details regarding the two numerical codes that we use. 

\subsubsection{Lagrangian code -- V1D}
\label{sec:V1D}

We use a modified version of the 1D, Lagrangian version of the VULCAN code \citep[hereafter V1D, for details, see][]{Livne1993IMT}, which solves the equations of reactive hydrodynamics. We modified V1D to be compatible with the input physics of Section~\ref{sec:input}. Linear artificial viscosity is only used for the HeCO setup (see details in Section~\ref{sec:HeCO results}). The Courant time step factor is $0.25$, and the maximum relative change of the density in each cell during a time step is smaller than $0.01$. Burning is not allowed on shocks (identified as cells with $q_{v}/P>0.1$, where $q_{v}$ is the artificial viscosity and $P$ is the pressure.). A simple semi-implicit Euler solver with adaptive time steps is used for the integration during the burning step. We choose a burning time step equals to the hydrodynamic time step, $\Delta t_{B}=\Delta t$, and iterate with the convergence criterion $\max_{i}(\Delta X_{i})<\delta_B$, where $\Delta X_{i}$ is the change in the composition $X_{i}$ over the last iteration, and $\delta_B=10^{-8}$. If the iterations do not converge, we decrease the burning time step and retry to solve. Following a successful iteration procedure, the burning time step is increased. The process ends when integration along the full $\Delta t$ has been completed. Initially, we divide the cells with equal size, $\Delta x_{0}$. The boundary condition for the left-hand boundary is a solid wall, and for the right-hand boundary it is a piston moving with the instantaneous velocity of the upstream plasma. The size of the computed domain is large enough to ensure that the leading shock do not reach the boundary before ignition is obtained. 

\subsubsection{Eulerian code -- FLASH}
\label{sec:FLASH}

We use a modified version of the Eulerian, 1D hydrodynamic FLASH4.0 code with thermonuclear burning \citep[][]{Fryxell2000,dubey2009flash}. We modified FLASH to be compatible with the input physics of Section~\ref{sec:input}. Instead of using the supplied burning routines of FLASH, which only support hard-wired $\alpha$-nets, we use the burning routines of V1D, including the same integration method. Specifically, instead of using one of the two integration methods supplied with FLASH (either fourth-order Rosenbrock method or variable-order Bader--Deuflhard method), we use the much simpler integration scheme of V1D. We find no significant difference between the simple V1D integration scheme and the fourth-order Rosenbrock method in a few cases. 

The simulation are performed in planar geometry, the cut-off value for composition mass fraction is $\textsc{smallx}=10^{-25}$, the Courant time step factor is $\textsc{CFL}=0.2$. Burning is not allowed on shocks and the nuclear burning time step factor is $\textsc{enucDtFactor}=0.2$. We divide the cells with equal size, $\Delta x$, which remain constant throughout the simulation (we do not use adaptive mesh refinement). This allows an easy interpretation of our results, with a price of longer simulation time that is acceptable for our 1D simulations. The boundary condition for the left-hand boundary is "reflected" (a solid wall). We override in each time step any deviations from the initial conditions because of the waves that develop next to the right-hand boundary, such that the shock always meets the initial upstream conditions. The size of the computed domain is large enough to ensure that the leading shock does not reach the boundary before ignition is obtained.  


\section{Ignition in a pure CO composition}
\label{sec:CO results}

\citet{Katz2019} claim that this setup for CO does not ignite, but this is because they arbitrarily stopped the simulation after $3.5\,\textrm{s}$\footnote{M. Katz, private communication.}, roughly $0.2\,\textrm{s}$ before ignition is obtained in our simulations\footnote{Also verified by M. Katz with their numerical code.}. 

We calculate this setup with both V1D and FLASH, and obtain a converged and resolved ignition. We preform simulations with and without a burning limiter \citep{Kushnir2013} and obtain similar results. The results shown below employ the limiter as described in \citet{Kushnir2013}, with the energy release per cell sound crossing time limited to a fraction $f=0.1$ of the thermal energy in the cell (when more energy is released, all rates are normalized by a constant factor to limit it). As explained below, this burning limiter suppresses artificial ignitions in low resolutions, but does not affect the process of ignition when it is resolved, as is the case here (see also Section~\ref{sec:discussion}, in which relevant comments by \citet{Katz2019} are addressed). 

\begin{figure}
\includegraphics[width=0.48\textwidth]{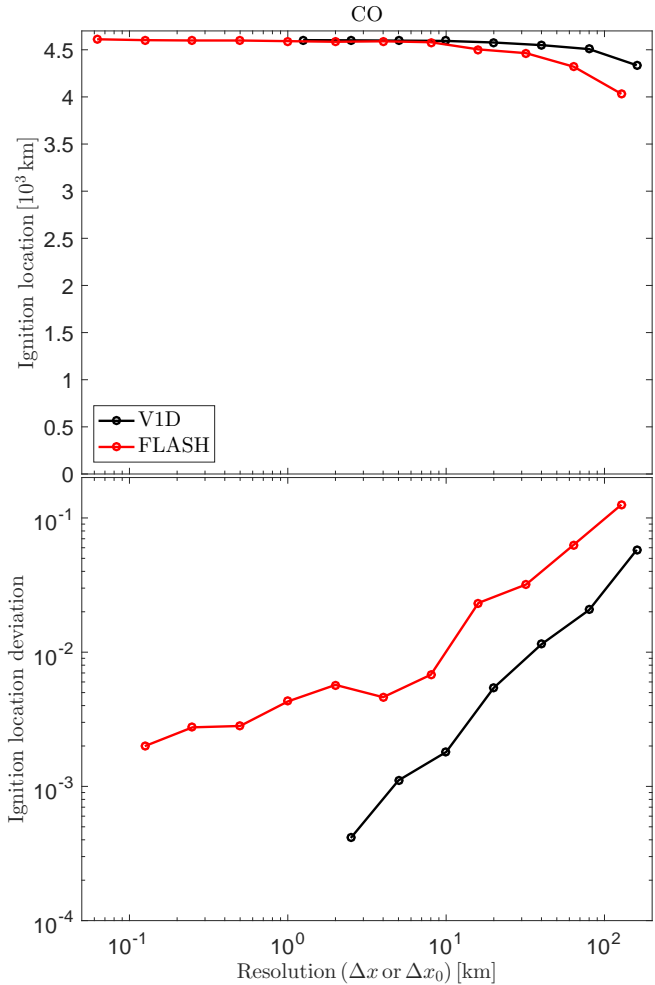}
\caption{A convergence study of the ignition location for pure CO mixture. V1D results are in black and FLASH results are in red. Top panel: the ignition location. Bottom panel: the deviation of the ignition location from the highest resolution result. Note that the plasma at the ignition location is compressed by a factor of $\mysim3$, such that the required resolutions for convergence are similar in both codes.
\label{fig:Uniform_Ign_CO_5e6}}
\end{figure}

An ignition of a detonation wave (two waves moving in opposite directions) is obtained at $x\simeq 4.45\times10^{3}\,\textrm{km}$ in both codes. A convergence study of the ignition location is shown in Figure~\ref{fig:Uniform_Ign_CO_5e6}. We use the simple criterion for the ignition location suggested by \citet{Katz2019}  -- the first point to satisfy $T_9>4$, which we find is a reasonable tracer of the ignition location for this test problem\footnote{This is, however, not the case for WD--WD collisions, see figure 1 of \citet{Kushnir2013}, where the contact region between the two stars reaches $T_9>7$ without leading to an ignition.}.  The V1D (FLASH) results are converged to better than $1\%$ for $\Delta x_{0}\lesssim 35\,\textrm{km}$ ($\Delta x\lesssim 10\,\textrm{km}$). Note that the plasma at the ignition location is compressed by a factor of $\mysim3$, such that the required resolutions for convergence are similar in both codes. The deviation between the converged locations of the two codes is $\mysim2\times10^{-3}$, consistent with the accuracy for which this location is determined. 

\begin{figure}
\includegraphics[width=0.48\textwidth]{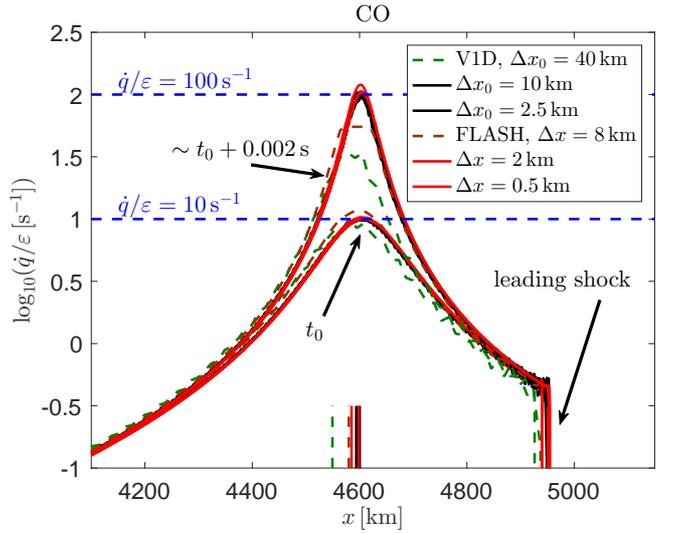}
\caption{A resolved ignition in the CO test problem. Snapshots of the relative burning rate, $\dot{q}/\varepsilon$, are presented in the series of the V1D (green and black) and FLASH (brown and red) simulations with increasing resolutions, at two times separated by about $2$ ms around $t_0\approx 3.7\,\textrm{s}$ after the beginning of the simulation. The profiles with high enough resolution are converged to high accuracy, such that the black and the red lines overlap. Note that the plasma at the ignition location is compressed by a factor of $\mysim3$ relative to $t=0$. The (initial) resolutions of the Lagrangian V1D code shown are thus larger to allow better comparison. The location at which the condition $T_9=4$ (shown in Figure~\ref{fig:Uniform_Ign_CO_5e6}) is shown as a short vertical line at the bottom of the plot for each resolution with corresponding color.
\label{fig:Uniform_Ign_hotspot_CO_5e6}}
\end{figure}

Ignition occurs robustly due to the shortening burning times behind the accelerating hydrodynamic shock, as described in \citep{Kushnir2013}. A demonstration that the ignition region is resolved is provided in Figure~\ref{fig:Uniform_Ign_hotspot_CO_5e6}. The burning rate $\dot{q}/\varepsilon$, where $\dot{q}$ is the energy injection from burning and $\varepsilon$ is the thermal energy, is shown in the vicinity of the ignition location, at two snapshots separated by about $0.002\,\textrm{s}$ around the onset of ignition. The speed of sound in this region is $c_s\approx 4\times 10^{3} \,\textrm{km}\,\textrm{s}^{-1}$. As can be seen, by the time of the second snapshot, a resolved region with a width of $\Delta x\sim 200 \,\textrm{km}$ and sound crossing time of $\mysim 0.05\,\textrm{s}$ is producing energy at a rate above $10\,\textrm{s}^{-1}$, which more than doubles within a time-scale of $0.002\,\textrm{s}$. A significant amount of energy is released within $\mysim0.01\,\textrm{s}$ and sound waves do not have sufficient time to distribute the excess pressure, resulting in two detonation fronts that form a short time later (not shown here). As can be seen, conditions for ignition \citep[significant energy release within less than a sound crossing time;][]{zeldovich,Kushnir2015} are obtained in a region that is well resolved and converged for resolutions higher than $\Delta x\sim 10\,\textrm{km}$. The burning limiter is not triggered in the first snapshot in figure \ref{fig:Uniform_Ign_hotspot_CO_5e6} at any of the resolutions shown, and is triggered in the second snapshot only in the lowest resolution, as apparent by the "flat-top" profiles in which the burning was limited. Note that for resolutions of few km, the ignition conditions are reached before the burning limiter is triggered. Even in the lowest resolutions where the limiter has the largest effect, the energy release is sufficient for igniting detonation waves. In fact, ignition at the same location was obtained in runs where the burning limiter was turned off across the simulation. 

As the temperatures continue to increase, the rate of burning increases and the scale on which burning occurs decreases substantially, leading to the well-known small length scale of thermonuclear detonation waves that cannot be resolved \citep[][]{Khokhlov89}. However, resolving this small length scale is no longer essential, as the downstream conditions are set by the total energy release \citep[][contrary to what is routinely iterated in the supernova literature, including by \citet{Katz2019}]{Kushnir2013,Kushnir2015}. 


\section{Ignition in a composition including $5\%$ He}
\label{sec:HeCO results}

\begin{figure}
\includegraphics[width=0.48\textwidth]{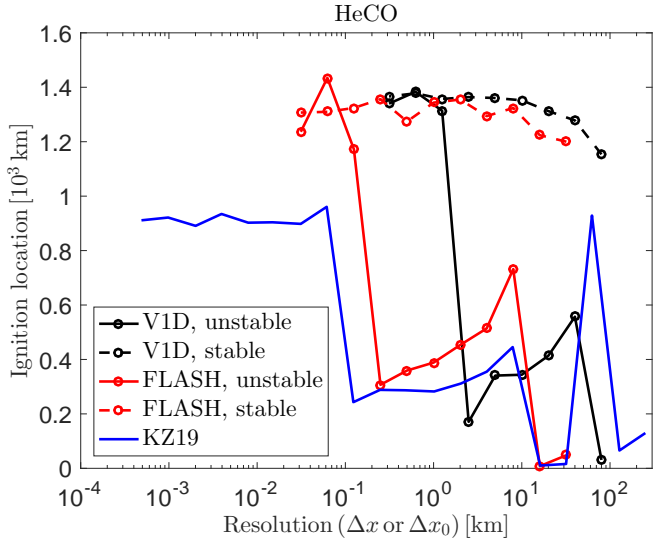}
\caption{A convergence study of the ignition location for the HeCO. The location is set in the same way as in \citep{Katz2019}, by identifying the first cell to reach a temperature of $T_9=4$. FLASH (V1D) results for simulations with the unstable boundary conditions are shown in solid red (black). Results for simulations with stable boundary conditions, where the burning was turned off in the first cell (three first cells for V1D) are shown in dashed lines with corresponding colors. The results of \citet{Katz2019} for the (relevant) case where a burning limiter is applied (their figure 4) are in solid blue. 
\label{fig:Uniform_Ign_He_5e6}}
\end{figure}

\citet{Katz2019} did achieve an ignition in a test problem where helium is included in the composition ($0.05$ by mass, leaving $0.45$ mass fraction for the oxygen, HeCO). They obtained ignitions at different locations for different resolutions regimes. Ignition was obtained close to the boundary for resolutions  $\Delta x \gtrsim 0.1\,\textrm{km}$, and at around $1000\,\textrm{km}$ for very high resolutions $\Delta x \lesssim 0.1\,\textrm{km}$. We calculate the evolution in a similar setup and obtain similar results as seen in solid lines in Figure~\ref{fig:Uniform_Ign_He_5e6}.  Note that there is a factor of $\mysim2$ difference in the required resolution for suppression of the artificial ignition in our results compared to that of \citet{Katz2019}, and that there is a different location of the ignition position.  These differences could be because of the different numerical schemes, or because of the different reaction networks ($178$ isotopes in our runs, compared with the $13$ isotope $\alpha$-net used by \citet{Katz2019}). In what follows, we assume that the basic issue that causes false ignition at low resolutions in our reproduced runs is the same as that in \citet{Katz2019}, although we are unable to verify this. We note that in addition to the problem with reaching convergence in the ignition location, we obtain strong fluctuations behind the shock (see Figure~\ref{fig:Uniform_Ign_hotspot_HeCO_5e6_no1} and discussion that follows). In the V1D runs, the strong fluctuations are suppressed using a small linear viscosity, with $q_{\textrm{lin}}=0.1\rho c_{s}|\Delta u|$, where $\rho$ is the density of the cell and $\Delta u$ is the difference of the velocity between the two nodes of the cell.

\begin{figure}
\includegraphics[width=0.48\textwidth]{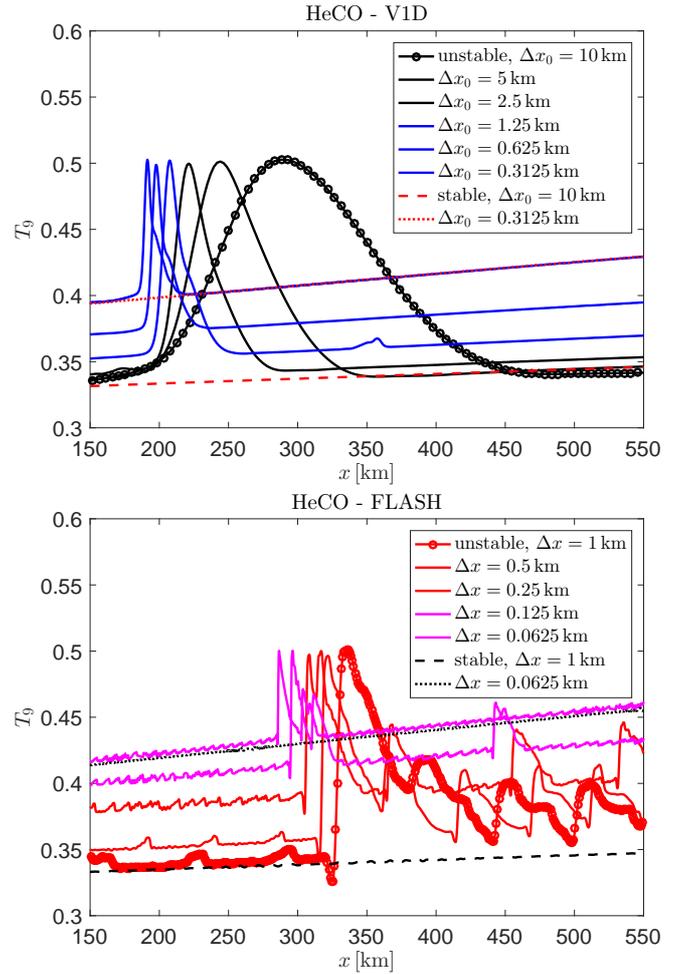}
\caption{The artificial hotspot region for HeCO mixtures. Snapshots of $T_{9}$ are presented in the series of the V1D (solid black and blue) and FLASH (solid red and magenta) simulations with the unstable boundary condition and increasing resolutions, when the maximum temperature of the hotspot reaches $T_9\approx0.5$ ($t\approx 0.6-1.1\,\textrm{s}$). Numerical ignition was obtained in simulations presented in solid black and solid red. Numerical ignition was not obtained in the higher resolution runs (blue and magenta) due to the small width of the hotspot. Note that a higher resolution is required for FLASH to suppress the artificial ignition (even after taking into account the factor of $\mysim3$ compression at the location of the artificial hotspot). Results for the lowest and highest resolutions, where burning was suppressed in the first few cells near the boundary, are shown at the corresponding times in dashed and dotted lines. 
\label{fig:Uniform_Ign_hotspot_HeCO_5e6}}
\end{figure}

\begin{figure}
\includegraphics[width=0.48\textwidth]{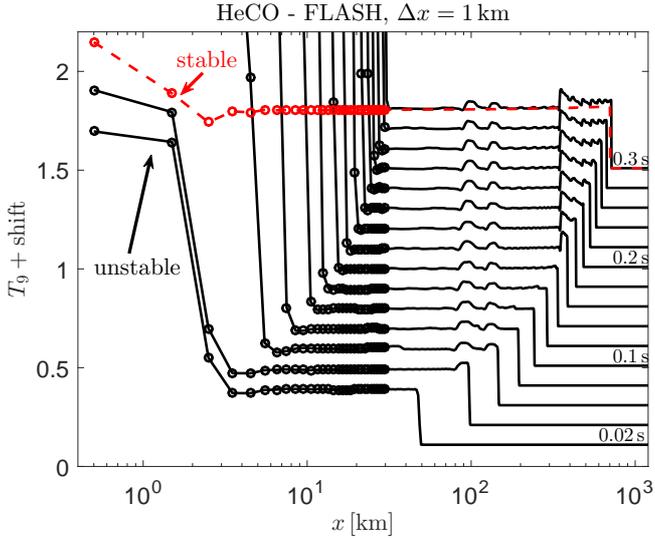}
\caption{The formation of the artificial hotspot for the HeCO mixture in the FLASH run with $\Delta x=1\,\textrm{km}$. Snapshots of $T_{9}$ are presented for the simulation with the unstable boundary condition (solid black) in increasing times, $t$, separated by $0.02\,\textrm{s}$. The profiles are shifted by $0.05(t/0.01\,\textrm{s})$ to enhance the visibility. The profile obtained at $t=0.3\,\textrm{s}$ from the run where burning was suppressed in the first cell near the boundary is shown in dashed red line. The first $30$ cells are marked with circles. Note the temperature in these cells does not reach the threshold $T_9=4$.
\label{fig:Early_He_5e6}}
\end{figure}

\begin{figure}
\includegraphics[width=0.48\textwidth]{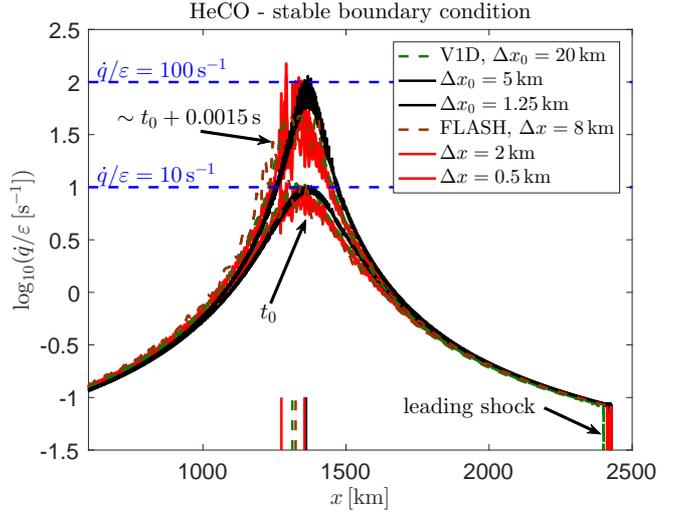}
\caption{The hotspot for the HeCO mixture with stable boundary conditions. Snapshots of $\dot{q}/\varepsilon$ are presented in the series of the V1D (green and black) and FLASH (brown and red) simulations with increasing resolutions, at two times separated by about $1.5$ ms around $t_0\approx 1.8\,\textrm{s}$ after the beginning of the simulation. Up to the numerical noise, the V1D (FLASH) profiles with high enough resolution are converged, such that the black (red) lines overlap. Note that the plasma at the ignition location is compressed by a factor of $\mysim2$, such that similar resolutions are compared for both codes. The location at which the condition $T_9=4$ (shown in Figure~\ref{fig:Uniform_Ign_He_5e6}) is shown as a short vertical line at the bottom of the plot for each resolution with corresponding color.
\label{fig:Uniform_Ign_hotspot_HeCO_5e6_no1}}
\end{figure}

The vicinity of the location of the false ignition that occurs in low resolutions is shown in Figure~\ref{fig:Uniform_Ign_hotspot_HeCO_5e6} for different resolutions. As can be seen, a hotspot is formed a few hundred $\rm km$ from the boundary with an unconverged profile that becomes smaller with better resolutions. The formation of this hotspot is demonstrated for the FLASH run with $\Delta x=1\,\textrm{km}$ in Figure~\ref{fig:Early_He_5e6}. The first cell near the boundary is numerically heated to a temperature above the exact solution of the colliding slab. This is a common deficit of both Lagrangian and Eulerian schemes near reflecting boundaries, which usually have a small effect. However, because of the fast reaction $^{16}$O$+\alpha$, the excess temperature is enough to burn the first cell and send a weak sound wave into the plasma. The hot plasma from the burnt cell is advected into the yet unburnt cell next to it\footnote{This 'mixing problem' is described in detail by \citet{Glasner2018}, along with a few suggestions to suppress it.} (the velocity of the cells near the boundary is negative), heating it up, and causing it to burn and to send another sound wave. This process repeats itself, sending more and more sound waves into the plasma. In fact, each oscillation seen in the bottom panel of Figure~\ref{fig:Uniform_Ign_hotspot_HeCO_5e6} is due to the burning of one additional cell near the boundary. The interaction of these sound waves in some specific locations amplifies the temperature perturbation, which finally leads to the creation of false hotspots (see several bumps in Figure~\ref{fig:Early_He_5e6}). Note that the FLASH temperature profiles in Figure~\ref{fig:Uniform_Ign_hotspot_HeCO_5e6} are more noisy than the V1D results, and that higher resolution is required for FLASH to suppress the artificial ignition (even after taking into account the factor of $\mysim3$ compression at the location of the artificial hotspot, Figure~\ref{fig:Uniform_Ign_He_5e6}). The reason is that more cells are burned near the boundary in the Eulerian case, because of the advection problem discussed above.

Numerical issues often occur near the boundary and require special treatment. In this case, a simple solution is to suppress burning in the first three cells (one cell) near the boundary for the V1D (FLASH) runs. As can be seen in the dashed and dotted lines in Figures~\ref{fig:Uniform_Ign_hotspot_HeCO_5e6} and~\ref{fig:Early_He_5e6}, the artificial hotspots are not produced once the artificial burning in the cells at the boundary is suppressed. We note that one may consider other ways to suppress the artificial burning, such as to start the simulation at some small time t>0 with an analytical solution for this time\footnote{In this case, the analytical solution in the downstream for $t\ll v_0/g_0$ (with $v_0=-2\times10^{8}\,\textrm{cm}\,\textrm{s}^{-1}$ and $g_0=-1.1\times10^{8}\,\textrm{cm}\,\textrm{s}^{-2}$) is given by $T_9\approx0.286$, a density of $\approx9.06\times10^{6}\,\textrm{g}\,\textrm{cm}^{-3}$, and an adiabatic index of $\approx1.49$. The shock Mach number is $\approx1.49$ and the velocity of the shock is $\approx2.47\times10^{8}\,\textrm{cm}\,\textrm{s}^{-1}$ (in the downstream frame).}. Such a method could be useful for simulating two slabs colliding with each other (without a reflecting boundary), where a similar artificial burning will take place away from the boundary.

As can be seen in Figure~\ref{fig:Uniform_Ign_hotspot_HeCO_5e6_no1}, with the stable boundary conditions a resolved ignition region with a width of $\mysim 200\,\textrm{km}$ forms similarly to the CO results. In this HeCO case, however, significant fluctuations exist in the ignition region (especially in the FLASH calculations) even when the stable boundary conditions are used. These are due to an enhancement of the post-shock numerical oscillations due to the fast $^{16}$O$+\alpha$ reaction. Despite the fluctuations, converged ignition is obtained in both codes at consistent locations (Figure~\ref{fig:Uniform_Ign_He_5e6}). Admittedly, the robustness of the ignition in this case is harder to demonstrate than for the pure CO case. We note that the fluctuations are greatly reduced in compositions with less He -- a modest reduction of He from $5\%$ to $4\%$ significantly reduces the fluctuations. On the other hand, higher He fractions increase the fluctuations and require care.

\section{Summary}
\label{sec:discussion}

Contrary to the claim by \citet{Katz2019}, ignition is obtained in a pure CO composition for the 1D test problem that they studied without the need to artificially increase the temperature or to add substantial amounts of He. This is demonstrated using Eulerian (FLASH) and Lagrangian (V1D) codes that obtain consistent ignition locations to within $\mysim 2\times10^{-3}$ (see Figure~\ref{fig:Uniform_Ign_CO_5e6}). Furthermore, the ignition region has a size of $\mysim100\,\textrm{km}$ and is resolved by simulations with resolutions of $\Delta x\lesssim 10\,\textrm{km}$ \citep[see Figure~\ref{fig:Uniform_Ign_hotspot_CO_5e6}, consistent with the original claim of][]{Kushnir2013}. Unlike the claim by \citet{Katz2019}, high resolutions that are much finer than $1\,\textrm{km}$ are not needed and produce identical results to lower converged resolutions. The fact that \citet{Katz2019} did not obtain an ignition for this setup is because they arbitrarily stopped the simulation after $3.5\,\textrm{s}$\footnote{M. Katz, private communication.}, roughly $0.2\,\textrm{s}$ before ignition is obtained in our simulations\footnote{Also verified by M. Katz with their numerical code.}.

We reproduce the convergence problem shown by \cite{Katz2019} for a composition that includes $5\%$ He (HeCO, see Figure~\ref{fig:Uniform_Ign_He_5e6}), and show that in our simulations, it is a result of unstable numerical burning in the cells adjacent to the boundary, which emits sound waves that later lead to a numerical ignition (see Figures~\ref{fig:Uniform_Ign_hotspot_HeCO_5e6} and~\ref{fig:Early_He_5e6}). It is likely that that the slow convergence obtained by \citet{Katz2019} in this case is because of a similar boundary numerical artifact, but we are unable to verify this. This problem can be easily avoided by suppressing the nuclear burning in the first one (three) cells next to the boundary for FLASH (V1D), reaching a converged ignition location (to about $10\%$) at resolutions $\Delta x\sim 10\,\textrm{km}$ in both the Eulerian and Lagrangian codes (see dashed lines in Figures~\ref{fig:Uniform_Ign_He_5e6} and~\ref{fig:Uniform_Ign_hotspot_HeCO_5e6_no1}). 

As noted by \citet{Katz2019}, the burning limiter \citep[see Section~\ref{sec:CO results} and][]{Kushnir2013} does not help to avoid the false ignition in this case. While the limiter is crucial for avoiding false ignition in realistic density profiles of WDs where the temperature diverges in the low-density contact region \citep{Kushnir2013,Kushnir2014}, it does not cure all possible boundary-related numerical ignitions. As argued by \citet{Katz2019}, a real physical ignition can be missed in simulations that use low resolutions and employ the limiter. However, contrary to \citet{Katz2019} claims, the use of the limiter is not to obtain converged results with low resolutions that do not resolve the ignition region. The ignition region in WD--WD collisions has a width of several tens of km (see Figures~\ref{fig:Uniform_Ign_hotspot_CO_5e6} and~\ref{fig:Uniform_Ign_hotspot_HeCO_5e6_no1}) and can be resolved with resolutions of few km. In such resolved ignitions, where significant energy is released within the sound crossing time of the region that spans many cells, the limiter will not suppress the ignition. Moreover, in such cases, the effect of the limiter will be smaller with higher resolution and convergence can be demonstrated. After ignition, as the temperature continues to rise, the burning length scales become too small to resolve. At this point, however, a detonation wave is inevitable and resolving these scales is not essential anymore. The threshold for ignition of $T_9=4$ applied by \citet{Katz2019} is obtained at such late times. Their figure 3 seems to indicate that the fast burning times require resolutions of $<10^{-3}\,\textrm{km}$ to resolve and is misleading. In reality, the $\sim100\,\textrm{km}$ wide ignition region occurs at lower temperatures with burning times that are orders of magnitude longer.

The main difference between the CO and the HeCO case is the presence of the fast reaction $^{16}$O$+\alpha$. While the resulting unstable boundary condition in the HeCO can (and should) be easily fixed, this reaction produces in addition large temperature fluctuations behind the shock (see Figure~\ref{fig:Uniform_Ign_hotspot_HeCO_5e6_no1}) that limit the accuracy of the convergence (few percents accuracy compared to sub-percent in the CO case). We note that these fluctuations strongly depend on the fraction of He. Even for a slightly smaller fraction of $4\%$, the fluctuations are significantly reduced and the convergence is better behaved. Accurate results are challenging for higher values of He.

Perhaps the most worrying claim by \citet{Katz2019} is that simulations with resolutions above $1~\rm km$ of the HeCO could be wrongly interpreted as converged ignition given that the ignition location seems to converge (see Figure~\ref{fig:Uniform_Ign_He_5e6}). A quick examination of the ignition region in these runs, however, reveals that the ignition is clearly not converged (Figure~\ref{fig:Uniform_Ign_hotspot_HeCO_5e6}). While we agree with \citet{Katz2019} that numerical simulations should not be trusted blindly, we believe that a physical understanding of the ignition process and a detailed analysis of the evolution are the way to proceed rather than a blind increase of the resolution.

\section*{Acknowledgements}
We thank Max Katz for useful discussions. DK is supported by the Israel Atomic Energy Commission -- The Council for Higher Education -- Pazi Foundation -- and by a research grant from The Abramson Family Center for Young Scientists. BK is supported by the Israeli Centers Of Research Excellence (ICORE) Program (1829/12), the Beracha Foundation and funding from the MINERVA Stiftung with the funds from the BMBF of the Federal Republic of Germany. 






\bsp	
\label{lastpage}
\end{document}